\documentclass[twocolumn]{jpsj2}

\def\runtitle{
Out-of-equilibrium 
 Anderson model at high and low bias voltages 
}

\def\runauthor{Akira {\sc Oguri}}

\title{ 
Out-of-equilibrium 
Anderson model at high and low bias voltages 
}

\author{Akira {\sc Oguri}}

\inst{
          Department of Material Science,
          Osaka City University, 
          Sumiyoshi-ku, Osaka 558-8585,
          Japan 
}

\recdate{May 22, 2002}

\abst{
We study the high- and low-voltage properties 
of the out-of-equilibrium Anderson model for quantum dots,  
using a functional method in the Keldysh formalism. 
The Green's function at the impurity site 
can be regarded as a functional of a nonequilibrium distribution 
function $f_{\text{eff}}(\omega)$. 
The dependence of the Green's function 
on the bias voltage $V$ and temperature $T$ arises 
through $f_{\text{eff}}(\omega)$. 
From this behavior as a functional,  
it is shown that 
the nonequilibrium Green's function at $eV\to\infty$ 
is identical to the equilibrium one at $T\to\infty$. 
This correspondence holds when
the couplings of the dot and two leads, at the left and right, are equal. 
In the opposite limit, for small $eV$, 
the low-energy behavior of the Green's function  
can be described by the local Fermi-liquid theory up to terms 
of order $(eV)^2$. 
These results imply that 
the correlation effects due to the Coulomb interaction $U$ can be 
treated adiabatically in the two limits, at high and low bias voltages. 
}

\kword{Kondo effect, Anderson model, Quantum dot, Nonequilibrium,
 Keldysh formalism}

\begin{document}

\sloppy

\maketitle

\section{Introduction}
\label{sec:Introduction}

The Kondo effect in quantum dots 
is a very active field of current research, 
and recent experiments
\cite{Goldharber,Cronenwett,Simmel,VanDerWiel}
have been shown to be in qualitative agreements 
with early predictions.\cite{NgLee,GlatzmanRaikh,Kawabata} 
Theoretically, the equilibrium and 
linear-response properties of the Kondo system realized  
in the quantum dots have been well understood through 
those of magnetic alloys\cite{Hewson} 
and precise calculations  
with the numerical renormalization group method.\cite{Izumida} 
However, the nonequilibrium properties under 
a finite bias voltage $V$ have not yet been fully understood. 
It is a novel problem of the strongly correlated electron systems, 
and has been studied extensively.\cite{WM,HDW2,TheoriesII,KNG,ao10}

At small voltages $eV \ll T_K$  near the linear-response regime,
nonequilibrium properties at low energies, i.e., 
at $T \ll T_K$ and $\omega \ll T_K$,  
can be described by the local Fermi-liquid theory,\cite{Nozieres,Yamada}
where $T_K$ is the Kondo temperature. 
Specifically, the nonlinear-response of the current 
through the quantum dots has been calculated up to terms of order $V^3$ 
based on the Kondo model\cite{KNG} and Anderson model.\cite{ao10}
The coefficients can be expressed in terms 
the correlation functions defined with respect 
to the equilibrium ground state,\cite{ao10}
and in the electron-hole symmetric case 
the coefficients are determined 
by $T_K$ and the Wilson ratio $R$.
In this paper, 
we suggest a procedure to estimate $T_K$ and $R$ 
experimentally from the differential conductance 
near the unitarity limit.

The main purpose of this paper is to describe a relation 
between the effects of the bias voltage and temperature  
based on a property of the generating functional $Z_J$ for 
the Keldysh Green's function of the Anderson impurity.
We show that $Z_J$ can be regarded 
as a functional of a nonequilibrium distribution 
function $f_{\text{eff}}(\omega)$, 
and the dependence of $Z_J$ on $eV$ and $T$ arises 
through $f_{\text{eff}}(\omega)$. 
From this feature of the generating functional, 
it is deduced that 
the nonequilibrium Green's function at $eV\to\infty$ 
is identical to the equilibrium one at $T\to\infty$. 
This correspondence 
holds exactly when the couplings between the dot and two leads are equal.
Furthermore, the low-voltage Fermi-liquid behavior mentioned above 
can also be deduced from this property of $Z_J$.
As the functional of Luttinger and Ward\cite{LuttingerWard}
has played an important role in the usual Fermi-liquid theory, 
some of the nonequilibrium properties can be deduced 
from those of the generating functional.

Our results obtained in the two opposite limits of $eV$ imply 
that the Coulomb interaction $U$ can be treated adiabatically  
at both low ($eV \ll T_K$) and high ($U \ll eV$) bias voltages. 
Therefore, as one of the possibilities,
we could expect that  
the perturbation theory in $U$ works for all values of $eV$, 
although another possibility that 
a different phase of non-perturbative nature exists 
at intermediate values of $eV$ could not be ruled out.

\section{Keldysh Formalism for Anderson Model}
\label{sec:model}

\subsection{Green's functions}

We start with the single Anderson impurity connected 
to two reservoirs at the left ($L$) and right ($R$):
\begin{align}
H &  = \  H_1 \ + \  H_2 \;,
\label{eq:H}
\\
H_1 & =  
\sum_{\lambda=L,R} 
 \sum_{k\sigma} 
  \epsilon_{k \lambda}^{\phantom{0}}\,
         c^{\dagger}_{k \lambda \sigma} 
         c^{\phantom{\dagger}}_{k \lambda \sigma}
 + 
  \sum_{\sigma}
      E_d \,  n_{d\sigma}
\,,
\label{eq:H1}
\\
H_2 &  =
 \sum_{\lambda=L,R} 
 \sum_{\sigma} v_{\lambda}^{\phantom 0}  \left(\,  
             d^{\dagger}_{\sigma}  c^{\phantom{\dagger}}_{\lambda \sigma}
      +   c^{\dagger}_{\lambda \sigma}  d^{\phantom{\dagger}}_{\sigma}
             \,\right)   
  \, + \, U\, n_{d\uparrow}  n_{d\downarrow} \;,
\label{eq:H2}
\end{align}
where $d_{\sigma}$ annihilates an electron with spin 
$\sigma$ at the dot,
 and
$n_{d\sigma} = d^{\dagger}_{\sigma} d^{\phantom{\dagger}}_{\sigma}$.
In the leads $\lambda$ ($= L,\, R$),
the excitations are described 
by $\epsilon_{k \lambda} = \epsilon_{k}  + eV_{\lambda}$ and 
corresponding eigenfunction $\phi_{k \lambda}^{\phantom{0}}(r)$.
To specify the static potentials $V_{\lambda}$, 
we introduce additional parameters $\alpha_{\lambda}$ as  
$V_L = \alpha_L V$ and $V_R = - \alpha_R V$ with $\alpha_L + \alpha_R =1$. 
Here, the Fermi level at equilibrium is taken to be the origin of the energy.
The onsite potential $E_d$ is assumed to be 
a constant which does not depend of the bias voltage.  
The mixing matrix elements $v_{\lambda}$ describe  
the couplings between the dot and leads,
and $c_{\lambda \sigma}^{\phantom{\dagger}} 
= \sum_k c_{k \lambda \sigma}^{\phantom{\dagger}} 
\phi_{k \lambda}^{\phantom{0}}(r_\lambda^{\phantom{0}})$ annihilates
an electron at the interface $r_\lambda^{\phantom{0}}$.
We assume the local density 
of states $\rho_{\lambda}^{\phantom{0}}(\omega) 
= \sum_k 
\left|\phi_{k \lambda}^{\phantom{0}}(r_\lambda^{\phantom{0}})\right|^2 
\delta(\omega-\epsilon_{k \lambda}^{\phantom{0}})$  
to be a constant and its band width $D$ to be infinity.
We will use units $\hbar=1$.

\begin{figure}[t]
\begin{center}
\setlength{\unitlength}{0.65mm}
\begin{picture}(120,20)(-5,0) 
\thicklines

\put(100,10){\oval(12,9)[r]} 

\put(60,5.5){\vector(1,0){1}}
\put(58.5,14.5){\vector(-1,0){1}}

\put(20,14.5){\line(1,0){80}}
\put(20,5.5){\line(1,0){80}}

\put(7,14){\makebox(0,0)[bl]{$-\infty$}}
\put(7,4){\makebox(0,0)[bl]{$-\infty$}}
\put(110,9){\makebox(0,0)[bl]{$+\infty$}}

\put(66,-1){\makebox(0,0)[bl]{$-$ branch}}
\put(66,17){\makebox(0,0)[bl]{$+$ branch}}

\end{picture}
 
\caption{The Keldysh contour of the time evolution.}
\label{fig:contour}
\end{center}
\end{figure}
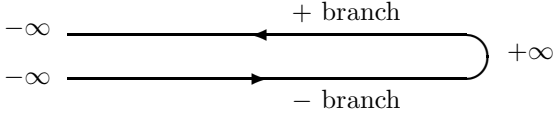

In order to describe a nonequilibrium steady state 
realized under the applied bias voltage, 
we employ the Keldysh formalism.\cite{KeldyshCaroliLandau} 
We define four types of local Green's functions 
at the impurity site;
\begin{align} 
 G^{--}(t) &= -\text{i} \,
\langle \text{T}\,
d^{\phantom{\dagger}}_{\sigma}(t)\, d^{\dagger}_{\sigma}(0)
\rangle  \;,
\\
G^{-+}(t) &=    \text{i}\, 
\langle 
d^{\dagger}_{\sigma}(0) \, d^{\phantom{\dagger}}_{\sigma}(t) 
\rangle \;, 
\\
G^{+-} (t) &=  - \text{i} \,
\langle 
d^{\phantom{\dagger}}_{\sigma}(t) \, 
d^{\dagger}_{\sigma}(0)  
\rangle \;,
\\
G^{++}(t) &= -\text{i} \,
\langle \widetilde{\text{T}}\,
d^{\phantom{\dagger}}_{\sigma}(t)\, d^{\dagger}_{\sigma}(0)
\rangle\;, 
\end{align} 
where
$\text{T}$ and $\widetilde{\text{T}}$ denote 
the time-ordering and anti-time-ordering operations, respectively,
and $d^{\phantom{\dagger}}_{\sigma}(t)$ is a Heisenberg operator. 
Note that these functions are linearly dependent 
 $G^{-+}+G^{+-}=G^{--}+G^{++}$,
and the retarded and advanced Green's functions 
are given by 
 $G^r = G^{--}-G^{-+}$ and $G^a = G^{--}-G^{+-}$. 
The average $\langle \cdots \rangle$ is 
taken over the density matrix $\widehat{\rho}(t)$ at $t=0$. 
The initial condition is given at $t=-\infty$,
where the two leads are separated from the dot 
and each part is in its own thermal 
equilibrium  described by $H_1$ and the chemical 
potentials $\mu_{\lambda} = eV_{\lambda}$. 
Then $H_2$, which includes  
the mixing $v_{\lambda}^{\phantom 0}$ and interaction $U$,
is switched on adiabatically.
The time evolution can be treated in
the interaction representation (see Appendix\ref{sec:time_evolution}), 
where the Wick's theorem is applicable for 
the time-ordered correlation functions 
along the Keldysh contour shown in Fig.\ \ref{fig:contour}.

\begin{figure}[t]
\begin{center}
\setlength{\unitlength}{0.6mm}
\begin{picture}(105,49)(-56,-10)
\thicklines

\put(-45,0){\vector(1,0){105}}
\put(0,0){\vector(0,1){40}}

\put(-45,25){\line(1,0){30}}
\put(-15,15){\line(0,1){10}}
\put(-15,15){\line(1,0){40}}
\put(25,0){\line(0,1){15}}

\put(0,-5){\makebox(0,0){\large $0$}}
\put(-15,-5){\makebox(0,0){\large $\mu_R^{\phantom{0}}$}}
\put(27,-5){\makebox(0,0){\large $\mu_L^{\phantom{0}}$}}
\put(54,-7){\makebox(0,0){\large $\omega$ }}
\put(11,37){\makebox(0,0){\large $f_{\text{eff}}(\omega)$}}
\put(-3,26){\makebox(0,0){\large $1$}}
\put(48,8){\makebox(0,0)
   {\Large $\frac{\mathstrut \Gamma_L}{\mathstrut \Gamma_L+\Gamma_R}$}}

\thinlines

\put(37,10){\vector(0,-1){10}}
\put(37,10){\vector(0,1){5}}

\end{picture}
\caption{
Nonequilibrium distribution $f_{\text{eff}}(\omega)$
at $T=0$.}
\label{fig:distribution}
\end{center}
\end{figure}
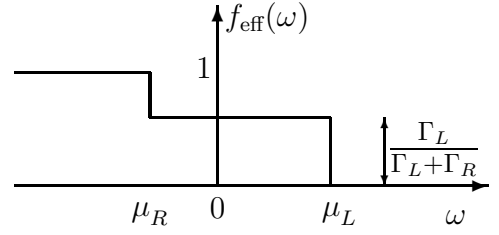

In the noninteracting case $U=0$, 
the Green's functions 
which include all contributions 
of the mixing $v_{\lambda}^{\phantom{0}}$ are given,  
as functions of the frequency $\omega$ (see 
Appendix\ref{sec:Matrix_Dyson}), by 
\begin{align}
 G_{0}^{--}(\omega) &=  
 \left[ 1-f_{\text{eff}}(\omega) \right]
     G_{0}^r(\omega)
     \, + \,  f_{\text{eff}}(\omega)\,G_{0}^a(\omega) 
\;,
\label{eq:G_0^--} 
\\
G_{0}^{-+}(\omega) &=  
-\,f_{\text{eff}}(\omega)
        \left[\,G_{0}^r(\omega)- G_{0}^a(\omega)\,\right] , 
\label{eq:G_0^-+}
\\
G_{0}^{+-}(\omega) &=  
\left[ 1-f_{\text{eff}}(\omega) \right]
        \left[\,G_{0}^r(\omega)- G_{0}^a(\omega)\,\right] , 
\label{eq:G_0^+-}
\\
 G_{0}^{++}(\omega) &=  
 -\left[ 1-f_{\text{eff}}(\omega) \right]
     G_{0}^a(\omega)
     \, - \,  f_{\text{eff}}(\omega)\,G_{0}^r(\omega) 
\;.
\label{eq:G_0^++}
\end{align}
Here $G_{0}^{r}(\omega) =  
  \left[\,\omega-E_d + \text{i}\, \Delta \,\right]^{-1}$ and
$G^a(\omega) = \{G^r(\omega)\}^*$.  
The level width is 
$\Delta = \Gamma_L + \Gamma_R$ with
 $\Gamma_{\lambda} = \pi \rho_{\lambda}^{\phantom{0}} v_{\lambda}^2$.
The function $f_{\text{eff}}(\omega)$ expresses 
the nonequilibrium distribution of the electrons 
at the impurity site,\cite{HDW2}
\begin{equation}
f_{\text{eff}}(\omega) =
{  f_L(\omega) \,\Gamma_L 
  + f_R(\omega) \,\Gamma_R
  \over 
 \Gamma_L +\Gamma_R } \;,
\label{eq:f_0}
\end{equation}
where  $f_{\lambda}(\omega) 
= f(\omega - \mu_{\lambda}^{\phantom{0}})$ 
with $f(\omega)=[\,e^{\omega/T}+1\,]^{-1}$.
At $T=0$,  $f_{\text{eff}}(\omega)$  has two steps 
at $\omega=\mu_L$ and $\mu_R$ as shown in Fig.\ \ref{fig:distribution}.
Note that the Keldysh formalism is applicable 
also in equilibrium, i.e., at $eV=0$, 
where $f_{\text{eff}}(\omega)$ becomes equal 
to the usual Fermi function $f(\omega)$.

The Green's function  for $U \neq 0$ satisfies  
the matrix Dyson equation;
$
\{\mbox{\boldmath $G$}(\omega)\}^{-1} 
    =     
 \{\mbox{\boldmath $G$}_{0}(\omega)\}^{-1} 
-  \mbox{\boldmath $\Sigma$}(\omega)\, 
$,
\begin{equation}
 \mbox{\boldmath $G$}_{0} 
\, = \, 
\left[\, 
 \begin{matrix}
  G^{--}_{0} & G^{-+}_{0}   \cr
  G^{+-}_{0} & G^{++}_{0}  \cr  
 \end{matrix} 
 \, \right] , 
\quad
\mbox{\boldmath $\Sigma$} \, = \, 
 \left[ \,
  \begin{matrix}
   \Sigma^{--} & \Sigma^{-+}  \cr
   \Sigma^{+-} & \Sigma^{++}  \cr 
  \end{matrix}
 \,\right] .
\label{eq:Keldysh_Matrix}
\end{equation}
Here $\mbox{\boldmath $\Sigma$}(\omega)$ is 
the self-energy due to the interaction $U$. 
Four types of the self-energies are also linearly 
dependent $\Sigma^{-+}+ \Sigma^{+-}= -\Sigma^{--}-\Sigma^{++}$. 
The perturbation theory 
for $\mbox{\boldmath $G$}(\omega)$ is 
described in the real-frequency (or real-time) representation,
and thus the dependence of $\mbox{\boldmath $G$}(\omega)$ 
on $eV$ and $T$ arises only 
through $f_{\text{eff}}(\omega)$ which enters the noninteracting 
one $\mbox{\boldmath $G$}_{0}(\omega)$. 
This feature seen in the Keldysh formalism is quite different from 
that of the Matsubara formalism  
in which the temperature dependence arises 
through the summations over the imaginary frequencies.

\subsection{Generating Functional}

In order to see this feature of the Keldysh formalism more explicitly,
we employ the generating functional $Z_J$ that yields 
the perturbation of series for $\mbox{\boldmath $G$}(\omega)$. 
It is given, in the path integral form,\cite{PathIntegral} by 
\begin{align}
Z_J  & \ \equiv \   
\int \!
{\cal D}\eta^{\dagger} {\cal D}\eta \  e^{\text{i} S(\eta^{\dagger},\eta)}
\;,
\label{eq:ZJ}
\\
   S(\eta^{\dagger}, \eta) & \ = \   
   S_0(\eta^{\dagger}, \eta) 
    \, + \,  
   S_{\text{ex}}(\eta^{\dagger}, \eta) 
      \, + \, 
   S_U(\eta^{\dagger}, \eta) \,. 
  \label{eq:action_tot}
\end{align}
The action $S$ consists of three parts 
corresponding to the free, external-source, and interaction contributions;  
\begin{align}
   S_0(\eta^{\dagger}, \eta) & \,=\,  
   \sum_{\sigma} \int_{-\infty}^{\infty} \! \text{d}t\,\text{d}t'\ 
 \mbox{\boldmath $\eta$}_{\sigma}^{\dagger}(t)\, 
       \mbox{\boldmath $K$}_0(t,t')\, 
 \mbox{\boldmath $\eta$}_{\sigma}(t')\,, 
 \label{eq:action_0} 
   \\
   S_{\text{ex}}(\eta^{\dagger},\eta) &\, = \, 
   -\sum_{\sigma} \! \int_{-\infty}^{\infty} \! \text{d}t \,
   \left[ \,
 \mbox{\boldmath $\eta$}_{\sigma}^{\dagger}(t) \,
       \mbox{\boldmath $J$}_{\sigma}(t)  +   
       \mbox{\boldmath $J$}_{\sigma}^{\dagger}(t) \, 
 \mbox{\boldmath $\eta$}_{\sigma}(t')
\,\right] ,
 \label{eq:action_ex} 
 \\
   S_U(\eta^{\dagger}, \eta) & \, = \, 
   -\,U \! \int_{-\infty}^{\infty} \! \text{d}t \,
 \Bigl[\,  
 \eta_{\uparrow -}^{\dagger}(t)\,
 \eta_{\uparrow -}^{\phantom{\dagger}}(t)\,
 \eta_{\downarrow -}^{\dagger}(t)\,
 \eta_{\downarrow -}^{\phantom{\dagger}}(t) \,
\nonumber \\
 & \qquad \;\;\; \quad -   \,
 \eta_{\uparrow +}^{\dagger}(t)\,
 \eta_{\uparrow +}^{\phantom{\dagger}}(t) \,
 \eta_{\downarrow +}^{\dagger}(t) \,
 \eta_{\downarrow +}^{\phantom{\dagger}}(t) \,\Bigr]  .
  \label{eq:action_U} 
\end{align}
Here $
 \mbox{\boldmath $\eta$}_{\sigma}^{\dagger}(t)
 = \left(\, 
 \eta_{\sigma -}^{\dagger}(t) \,, \, 
 \eta_{\sigma +}^{\dagger}(t) \,\right) 
 $ is a two component field of the Grassmann number. 
The label $\mp$ specifies the branches of the Keldysh contour 
(see Fig.\ \ref{fig:contour})  where 
each of the components corresponds to. 
The Kernel $\mbox{\boldmath $K$}_0$ in eq.\ (\ref{eq:action_0}) is 
the inverse matrix of the noninteracting Green's function, 
 \begin{align}
   \mbox{\boldmath $K$}_0(\omega)  & \ \equiv \ 
    \left\{\mbox{\boldmath $G$}_0(\omega)\right\}^{-1}
 \;, \\
    \mbox{\boldmath $K$}_0(t,t') &\  = \ 
    \int_{-\infty}^{\infty} \! {\text{d}\omega \over 2 \pi}\,
    \mbox{\boldmath $K$}_0(\omega) 
    \, e^{-\text{i}\omega (t-t')} 
        . 
 \end{align}
In eq.\ (\ref{eq:action_ex}), 
 $
 \mbox{\boldmath $J$}_{\sigma}^{\dagger}(t)
 = \left(\, 
 J_{\sigma -}^{\dagger}(t), \, 
 J_{\sigma +}^{\dagger}(t) \,\right) 
 $ is an external source of the anticommutating c-number, 
which is introduced for the later convenience.
In eq.\ (\ref{eq:action_U}), the sign of the first and second terms 
are determined by that of the exponent of the time-evolution operators 
${\cal U}(+\infty,-\infty)$ and ${\cal U}(-\infty,+\infty)$, respectively
(see Appendix\ref{sec:time_evolution}).
For $U=0$, the path integral of the fermionic fields 
in eq.\ (\ref{eq:ZJ}) can be evaluated analytically 
\begin{align}
Z_J^0 
&\, \equiv \, 
\int \!
{\cal D}\eta^{\dagger} {\cal D}\eta \  
e^{\text{i} \,\left[ S_0(\eta^{\dagger},\eta) 
+ S_{\text{ex}}(\eta^{\dagger},\eta) 
\right]}
\nonumber 
\\
&\, =\,  
Z_{J=0}^0
\,  \exp \! \left[ -\text{i}
 \sum_{\sigma} \! \int_{-\infty}^{\infty} \!\!\! \text{d}t\,\text{d}t'\,
 \mbox{\boldmath $J$}_{\sigma}^{\dagger}(t)\, 
       \mbox{\boldmath $G$}_0(t,t')\, 
 \mbox{\boldmath $J$}_{\sigma}(t') 
\right] ,
\label{eq:ZJ0}
\end{align}
where $Z_{J=0}^0 \equiv \lim_{J\to 0} Z_J^0$. 
Then following the standard prescription,\cite{PathIntegral} 
the generating functional and full Green's function can be rewritten 
in the form 
\begin{align}
Z_J &\ = \  e^{
\text{i} S_U\left(-\text{i} {\delta \over\delta J},\, 
\text{i} {\delta \over \delta J^{\dagger}}
  \right) } \ Z_J^0  \;,
  \label{eq:ZJwick}
  \\
G^{\nu\nu'}_{\sigma}(t,t') &\ = \ 
-\text{i}\,{1 \over Z_J} 
\left.
{\delta \over \delta J^{\dagger}_{\sigma\nu}(t)}\, 
{\delta \over \delta J^{\phantom{0}}_{\sigma\nu'}(t')}\, 
Z_J \,\right|_{J=0} ,
\label{eq:Green_path}
\end{align}
where $\nu$, $\nu'$ $= +,\,-\,$. 
In eq.\ (\ref{eq:ZJwick}), the fermionic fields 
in  $S_U$ have been replaced with the functional differentiations,  
$\eta^{\dagger}_{\sigma\nu}  \Rightarrow 
-\text{i}\, \delta / \delta J_{\sigma\nu} $ 
and 
$\eta^{\phantom{\dagger}}_{\sigma\nu}  \Rightarrow 
\text{i}\, \delta /\delta J^{\dagger}_{\sigma\nu}$.
The perturbation series of $\mbox{\boldmath $G$}$ can be 
generated from eq.\ (\ref{eq:Green_path}) by 
expanding $e^{\text{i}S_U}$ in eq.\ (\ref{eq:ZJwick}) as a power series 
of $S_U$ and then carrying out the functional differentiations 
using the explicit expression of $Z_J^0$ given by eq.\ (\ref{eq:ZJ0}). 
Each term of the perturbation series is written 
in terms of $\mbox{\boldmath $G$}_0$  
and the integrations over the internal variables 
of real time (or real frequency).  
Therefore, $Z_J$ can be regarded as a functional of $f_{\text{eff}}(\omega)$,
and the dependence of $\mbox{\boldmath $G$}(\omega)$ 
on $eV$ and $T$ arises only through $f_{\text{eff}}(\omega)$.
From this of property, the behavior 
of $\mbox{\boldmath $G$}(\omega)$  at high and low bias voltages 
can be deduced exactly. 


\section{High-Voltage Behavior}
\label{sec:high-voltage}

\subsection{General properties}

We consider the behavior of $\mbox{\boldmath $G$}(\omega)$ 
in the high-voltage limit $eV \to \infty$.
To be specific, we assume that the bias voltage 
is applied to be $\alpha_L>0$ and $\alpha_R>0$.
In the limit of $eV \to \infty$, 
the chemical potentials tend to $\mu_L \to \infty$ and $\mu_R \to -\infty$,
and thus the distribution function becomes a constant  
$\left. f_{\text{eff}}(\omega)\right|_{eV \to\infty} 
\equiv \Gamma_L/(\Gamma_L + \Gamma_R)$,  
see Fig.\ \ref{fig:distribution}.
When the couplings with the two leads are equal $\Gamma_L=\Gamma_R$,
the value of this constant becomes $1/2$  
and $\left. f_{\text{eff}}(\omega)\right|_{eV \to\infty}$ 
coincides with the high-temperature limit of the usual Fermi function 
 $\left. f(\omega)\right|_{T \to\infty} \equiv 1/2$.
Therefore, the noninteracting Green's 
function $\mbox{\boldmath $G$}_0$ for $eV \to \infty$ becomes 
identical to that for $T \to \infty$ at equilibrium.
The same correspondence holds also 
for the full Green's function $\mbox{\boldmath $G$}$ 
since it is determined by eqs.\ (\ref{eq:ZJ0})--(\ref{eq:Green_path}) 
for given $\mbox{\boldmath $G$}_0$. 
Thus in the case of $\Gamma_L=\Gamma_R$ the two limits, 
{\em i}) $eV \to \infty$ and {\em ii}) $T\to \infty$ keeping $eV=0$, 
are equivalent as far as the Green's function 
at the impurity site is concerned. 
Note that the temperature is not necessary to be kept at $T=0$ 
in the $eV\to\infty$ limit.
In these two equivalent limits, 
the Coulomb interaction $U$ may be treated adiabatically. 
This is because in thermal equilibrium the perturbation series in $U$ is 
absolutely convergent at $T=0$ for any finite $U$,\cite{ZlaticHorvatic}
and there is no phase transition at finite $T$.  
When the voltage is finite but still much larger than the other energy 
scales $eV \gg \max(U, |E_d|, \Delta, T)$, 
the behavior of $\mbox{\boldmath $G$}(\omega)$ at 
low frequencies $|\omega| \ll eV$ may be 
approximated reasonably by that in the $eV \to \infty$ limit.

In the case of $\Gamma_L \neq \Gamma_R$ the two limits,
{\em i}) and {\em ii}), are not equivalent because 
the charge distribution around the impurity at $eV\to \infty$ 
does not correspond to that at $T \to \infty$.
In this relation, we examine the $T \to \infty$ limit at finite $eV$.
In this limit the distribution function is given by 
$\left. f_{\text{eff}}(\omega)\right|_{T \to\infty} \equiv 1/2$
without the assumption of $\Gamma_L = \Gamma_R$, 
and the Green's function becomes identical to 
that in the $T\to \infty$ limit at $eV=0$. 
Physically, this is rather obvious because 
at $T \gg eV$ the effects of the bias voltage are hidden
by the thermal fluctuations.

\begin{figure}[t]
\leavevmode 
\begin{center}
\begin{minipage}{0.9\linewidth}
\includegraphics[width=\linewidth, clip, 
trim = 0cm 0.8cm 0cm 0cm]{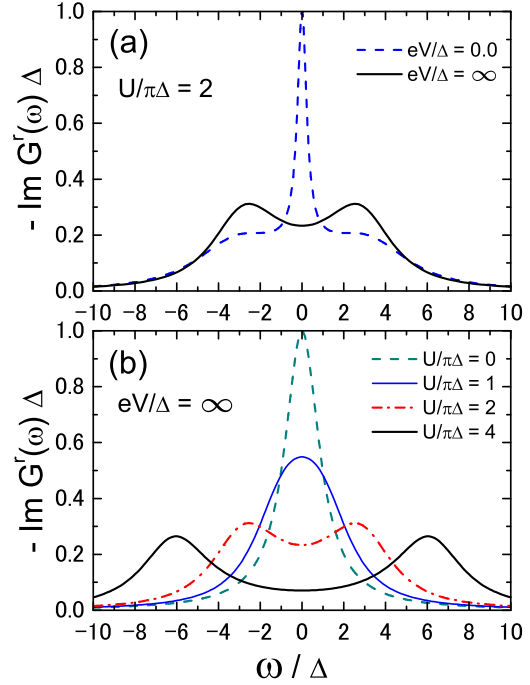}
\end{minipage}
\caption{Spectral function obtained with order $U^2$ self-energy.} 
\label{fig:G_inf}
\end{center}
\end{figure}

\subsection{Order $U^2$ self-energy}

In order to show a rough sketch of the Green's function 
in the high-voltage limit,
we use here the order $U^2$ retarded 
self-energy.\cite{Yamada,HDW2,HorvaticSokcevicZlatic} 
It yields a reliable picture at least qualitatively,
and can be calculated analytically at  $eV \to \infty$,
\begin{equation}
\left.
\Sigma^{r(2)}_{\text{cor}}(\omega)\right|_{V\to\infty} \ = \ 
   \left({U \over 2}\right)^2 { 1\over \omega\, + \,\text{i}\, 3\Delta} \;. 
\label{eq:self_U2}
\end{equation}
Here we have assumed the 
electron-hole symmetry taking the parameters 
to be $E_d=-U/2$, $\mu_L= eV/2$, $\mu_R = -eV/2$,
and $\Gamma_L=\Gamma_R$ ($=\Delta/2$),
so that the results are applicable equally to 
the $T\to\infty$ limit at equilibrium. 
Note that  eq.\ (\ref{eq:self_U2}) is 
the correlation part, which is obtained by separating  
the Hartree-Fock contribution $U/2$ 
as $\Sigma^r = U/2 + \Sigma^r_{\text{cor}}$,
and the retarded Green's function is given by
 $G^r(\omega) = 
 [\omega + \text{i}\, \Delta - \Sigma^r_{\text{cor}}(\omega)]^{-1}$. 
Fig.\ \ref{fig:G_inf} shows the spectral function,  
$\text{Im} G^r(\omega)$, obtained using the order $U^2$ self-energy.
The Kondo resonance situated at $\omega=0$, $eV=0$ (dashed line) 
disappears at high voltages (solid line), 
as seen in Fig.\ \ref{fig:G_inf}(a) for $U/(\pi\Delta) =2$. 
The results in the $eV \to \infty$ limit are 
plotted for several values of $U$  in Fig.\ \ref{fig:G_inf}(b).
For strong interactions $U \gg \Delta$ the spectral function 
has two peaks corresponding to the Hubbard bands 
at $\omega \simeq \pm U/2$, 
while for weak interactions $U \lesssim \Delta$ the contributions of 
the mixing dominate and it results in the single-peak structure.
The order $U^2$ results could be refined quantitatively 
by including the higher-order terms\cite{Logan} or 
by numerical methods.

\section{Low-Voltage Behavior}
\label{sec:low-voltage}

\subsection{Ward identity}

In the opposite limit, at small bias voltages $eV \ll T_K$, 
the low-energy properties can be described 
by the Fermi-liquid theory.\cite{Nozieres,Yamada} 
The proof has been provided in the previous paper 
using the Ward identity 
for the first and second derivatives  
of the self-energy with respect to $eV$.\cite{ao10} 
We describe here the outline briefly to emphasize 
the properties of  $\mbox{\boldmath $\Sigma$}(\omega)$ as 
 a functional of the distribution function $f_{\text{eff}}(\omega)$. 
Since the voltage $V$ enters $\mbox{\boldmath $G$}_0(\omega)$ 
through $f_{\text{eff}}(\omega)$ 
as seen in eqs.\ (\ref{eq:G_0^--})--(\ref{eq:f_0}),
we have    
\begin{align}
&
\!\!\!\!
\left.
{\partial  
\over
\partial ({\sl e}V)} \mbox{\boldmath $G$}_0(\omega) \right|_{V=0}
= \,
 -\, \alpha 
\left(
{\partial \over \partial \omega}
+ {\partial \over \partial E_d}
\right)
\mbox{\boldmath $G$}_{0:\text{eq}}(\omega)\,,
 \label{eq:derivative_1b}
\\
&
\!\!\!\!
\left.
{\partial^2  
\over
\partial ({\sl e}V)^{2}} \mbox{\boldmath $G$}_0(\omega)\right|_{V=0}
 = \,
\kappa 
\left(
{\partial \over \partial \omega}
+ {\partial \over \partial E_d}
\right)^{2}
\mbox{\boldmath $G$}_{0:\text{eq}}(\omega)\,.
 \label{eq:derivative_2b}
\end{align}
Here 
$
\mbox{\boldmath $G$}_{0:\text{eq}}(\omega) 
\equiv 
\left.\mbox{\boldmath $G$}_{0}(\omega)\right|_{V=0}$ 
is the equilibrium Green's function,
$\,\alpha \equiv (\alpha_L \Gamma_L - \alpha_R \Gamma_R)/ 
   (\Gamma_L+ \Gamma_R)$, and 
$\kappa \equiv 
    {
   (\alpha_L^2 \Gamma_L + \alpha_R^2 \Gamma_R)/ 
  (\Gamma_L+ \Gamma_R})$.
The derivatives of 
$\mbox{\boldmath $\Sigma$}(\omega)$ with respect to $eV$ can be 
calculated by taking the derivative 
of $\mbox{\boldmath $G$}_{0}$ in each term of the perturbation series,
and then replacing the derivative $\partial /\partial (eV)$ with 
$(\partial /\partial \omega + \partial/ \partial E_d )$ using 
eqs.\ (\ref{eq:derivative_1b}) and (\ref{eq:derivative_2b}).  It yields 
\begin{align}
\left.
{\partial  
\mbox{\boldmath $\Sigma$}(\omega) 
\over
\partial ({\sl e}V)}\right|_{V=0} 
 &= \, 
 -\,
 \alpha  
\left(
{\partial \over \partial \omega}
+ {\partial \over \partial E_d}
\right)
\mbox{\boldmath $\Sigma$}_{\text{eq}}(\omega) 
\;,
\label{eq:derivative_self_1}
\\
\left.
{\partial^2  
\mbox{\boldmath $\Sigma$}(\omega) 
\over
\partial ({\sl e}V)^2}\right|_{V=0} 
&= 
\, \alpha^2
\left(
{\partial \over \partial \omega}
 + {\partial \over \partial E_d}
\right)^2 
\mbox{\boldmath $\Sigma$}_{\text{eq}}(\omega) 
 \nonumber \\
 & \quad + \,    
{ \Gamma_L\,\Gamma_R 
  \over \left( \Gamma_L+ \Gamma_R \right)^2}
   \  \widehat{D}^2 
\mbox{\boldmath $\Sigma$}_{\text{eq}}(\omega) 
\;,
\label{eq:derivative_self_2_with_CII}
\end{align}
where 
$\mbox{\boldmath $\Sigma$}_{\text{eq}}(\omega) 
\equiv \left. \mbox{\boldmath $\Sigma$}(\omega) \right|_{V=0}$.   
In eq.\ (\ref{eq:derivative_self_2_with_CII}), 
$\widehat{D}^2$ denotes the functional operation 
carrying out the second derivative 
$(\partial /\partial \omega + \partial/ \partial E_d )^2$ 
for all the single $\mbox{\boldmath $G$}_0$'s in the perturbation series 
of $\mbox{\boldmath $\Sigma$}_{\text{eq}}$,  
which can formally be expressed using 
the functional differentiation of 
 $\mbox{\boldmath $\Sigma$}_{\text{eq}}$  with respect 
 to $\mbox{\boldmath $G$}_{0}$, 
\begin{align}
   \widehat{D}^2 
\mbox{\boldmath $\Sigma$}_{\text{eq}}(\omega)
 \, &\equiv\,   \sum_{\nu\nu'} 
\int  {\text{d}\omega'}  \, 
{
\delta \mbox{\boldmath $\Sigma$}_{\text{eq}}(\omega)
\over
\delta G_{0:\text{eq}}^{\nu\nu'}(\omega')
} 
 \nonumber
 \\
 & \qquad \qquad \times 
\,
\left(
{ \partial 
  \over
  \partial \omega' }
+{ \partial 
  \over
  \partial E_d }
  \right)^2   G_{0:\text{eq}}^{\nu\nu'}(\omega') \;.
\end{align}
Note that 
the perturbation series can be described in the real-frequency representation, 
and $\mbox{\boldmath $\Sigma$}(\omega)$ can be regarded as 
a functional of $\mbox{\boldmath $G$}_0(\omega)$. 
Eqs.\ (\ref{eq:derivative_self_1}) and (\ref{eq:derivative_self_2_with_CII})
relate the nonequilibrium quantities with the equilibrium ones. 
Especially, 
at $T=0$ the usual zero-temperature formalism 
of the Green's function is applicable,
and the right-hand side of 
eqs.\ (\ref{eq:derivative_self_1}) and (\ref{eq:derivative_self_2_with_CII})
can be rewritten in terms of the vertex corrections.\cite{ao10}
Along this line, 
the low-energy behavior of the self-energy  $\Sigma^r(\omega)$ and 
spectral function $A(\omega) \equiv -\text{Im}\, G^{r}(\omega)/\pi$ 
has been calculated exactly 
up to terms of order $\omega^2$, $T^2$, and $(eV)^2$.

\begin{figure}[t]
\leavevmode 
\begin{center}
\begin{minipage}{0.9\linewidth}
\includegraphics[width=1\linewidth, clip, 
trim = 2.9cm 17.35cm 3cm 2.2cm]{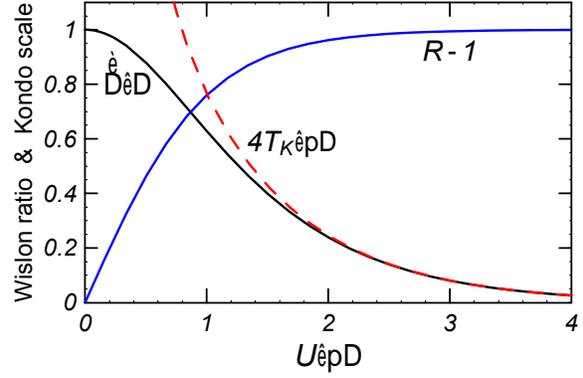}
\end{minipage}
\caption{The $U$ dependence of $R$ and $\widetilde{\Delta}\,$: 
Bethe ansatz results.} 
\label{fig:Bethe}
\end{center}
\end{figure}

\subsection{Experimental determination of the Wilson ratio}

The low-energy behavior of the differential conductance $dI/dV$ has also been 
calculated up to terms of order $T^2$ and $(eV)^2$ using   
the results of $A(\omega)$ and  
the formula for the nonequilibrium current eq.\ (\ref{eq:caroli}) 
given in Appendix\ref{sec:current}. 
Specifically, in the electron-hole symmetric case, 
the local Fermi liquid is characterized by two parameters, 
i.e., the Wilson ratio $R$ and the energy scale $\widetilde{\Delta}$ which 
corresponds to the width of the Kondo resonance.\cite{HewsonADV} 
In terms of these two parameters, the low-energy behavior of 
$A(\omega)$ and $dI/dV$ can be expressed in the form  
\begin{align}
& A(\omega)
\ = \ {1\over \pi \Delta}  
  \Biggl [\, 1 \,
  - \,\left( 1 + {(R-1)^2 \over 2} \right) 
  \left({\omega \over \widetilde{\Delta}}\right)^2
\nonumber \\
   & \qquad 
 \,
   -\, {(R-1)^2 \over 2}  
                 \left( {\pi T \over \widetilde{\Delta}}\right)^2 
     -\, {3\, (R-1)^2 \over 8} 
                 \left( {eV \over \widetilde{\Delta}}\right)^2 
 + \cdots  \,\Biggr] ,             
\label{eq:Spectral}
\\
& {dI \over dV}  
\ = \ {2 e^2 \over h}  
  \Biggl[\, 1 \,
  - 
                \, { 1
                  + 2\, (R-1)^2 
                  \over 3} 
                 \left( {\pi T \over \widetilde{\Delta}}\right)^2
   \nonumber \\
 &  \qquad \qquad \qquad \;
 \,
      - 
                \, { 1
                  + 5\, (R-1)^2 
                  \over 4} 
                 \left( {eV \over \widetilde{\Delta} } \right)^2 
 + \cdots 
 \,\Biggr] .              
\label{eq:dI_dV}
\end{align}
These parameters contain 
all contributions of the perturbation series in $U$,
and are defined with respect to the equilibrium ground state:
$R \equiv \widetilde{\chi}_s/\widetilde{\gamma}$ 
and $\widetilde{\Delta} \equiv \Delta/\widetilde{\gamma}$ 
where  $\widetilde{\gamma}$ and $\widetilde{\chi}_s$ are  
the enhancement factors for the $T$-linear specific heat 
and spin susceptibility, respectively.\cite{FermiLiquidRelations}
The values of these parameters can be evaluated using 
the exact Bethe ansatz solution,
\cite{KawakamiOkiji,WiegmanTsvelick,ZlaticHorvatic}
as shown in Fig.\ \ref{fig:Bethe}.
The width of the Kondo peak $\widetilde{\Delta}$ 
decreases monotonically with increasing $U$ and 
tends to $\widetilde{\Delta} \to 4 T_K /\pi$ for large $U$, where  
\begin{equation}
T_K \, = \, \pi \Delta \sqrt{u / (2 \pi)}\, \exp[-\pi^2 u/8 + 1/(2u)]  
\;,
\end{equation}
and $u= U/(\pi \Delta)$.
The Wilson ratio increases with $u$ from the noninteracting value $R=1$, 
and converges rapidly for $u \gtrsim 2.0$ to the value 
of the Kondo limit $R \to 2$ reflecting 
the suppression of the charge fluctuations 
by the strong Coulomb interaction. 
For the quantum dots, 
the value of $R$ and $\widetilde{\Delta}$ can 
be estimated experimentally from the results of $dI/dV$ 
without carrying out the measurements of the spin susceptibility 
and specific heat:  
the coefficient of the $T^2$ and $V^2$ terms  
of $dI/dV$ in eq.\ (\ref{eq:dI_dV}) may be estimated 
from the observations near the unitarity limit. 
This is also one typical feature of 
the Kondo system in the quantum dots.

\section{Summary}
\label{sec:SUMMARY}

Using the functional method, 
the dependence of the Keldysh Green's 
function $\mbox{\boldmath $G$}(\omega)$ on 
 $eV$ and $T$ has been confirmed to arise through the nonequilibrium 
 distribution function $f_{\text{eff}}(\omega)$ which 
 enters $\mbox{\boldmath $G$}_0(\omega)$. 
From this property, the asymptotic behavior of
 $\mbox{\boldmath $G$}(\omega)$  at high 
 and low bias voltages has been deduced exactly.
The low-voltage behavior is determined by 
a set parameters of the local Fermi liquid such as 
the Wilson ratio $R$ and Kondo energy scale $\widetilde{\Delta}$.
The values of these parameters can be estimated  
experimentally from the $T^2$ and $V^2$ contributions  
of the differential conductance near the unitarity limit.
In the high-voltage limit $eV\to\infty$ the 
Green's function becomes 
identical to that of the $T\to\infty$ limit at equilibrium, 
when the couplings between the dot 
and two leads are symmetric $\Gamma_L =\Gamma_R$.
These results suggest that the Coulomb interaction $U$ can be 
treated adiabatically at both low ($eV \ll T_K$) and 
high ($U \ll eV$) bias voltages.

\section*{Acknowledgements}

I would like to thank A. C. Hewson, H. Ishii and  
S. Nonoyama for discussions. 
This work has been supported by the Grant-in-Aid 
for Scientific Research from the Ministry of Education, 
Science and Culture, Japan.

\appendix

\section{Time Evolution of Density Matrix}
\label{sec:time_evolution}

The time evolution of the density matrix is described by 
\begin{equation} 
{\partial  \over \partial t} \, \widehat{\rho}(t) \ = \ -\text{i} \, 
\left[\, H \, , \, \widehat{\rho}(t) \, \right] 
\;.
\label{eq:Neumann}
\end{equation}
The formal solution of this equation 
can be obtained in the interaction representation,   
\begin{align} 
 \widetilde{\rho}(t) \,  &\equiv \,  
        e^{\text{i} H_1 t} \, \widehat{\rho}(t)\, e^{-\text{i} H_1 t}  
\nonumber
   \\      
  &= \, {\cal U}(t,t_0) \, 
                         \widetilde{\rho}(t_0) \, 
                         {\cal U}(t_0,t) 
\label{eq:rho} \;,
\end{align}
where ${\cal U}(t,t_0)$ is the time-evolution operator 
\begin{equation}
{\cal U}(t,t_0) \, = \, \text{T} \exp 
        \left[\,- \text{i} \int_{t_0}^t \text{d}t' \, 
        \widetilde{H}_2 (t') \,\right] 
\label{eq:U_2} 
\end{equation}
and $\widetilde{H}_2(t)  =  
      e^{\text{i} H_1 t}\, H_2\, e^{-\text{i} H_1 t}$.
The relations among the Schr\"odinger  ${\cal O}_S$,
interaction $\widetilde{\cal O}(t)$,
and Heisenberg ${\cal O}_H(t)$ operators are 
\begin{align}
\widetilde{\cal O}(t) &= 
 e^{\text{i} H_1t}\, {\cal O}_S \,e^{-\text{i} H_1 t} \;,
  \\
{\cal O}_H(t) &= {\cal U}(0,t) \, \widetilde{\cal O}(t) \, {\cal U}(t,0) \;.
\end{align}
To describe the nonequilibrium state, 
the initial condition is given at $t_0 \to -\infty$ 
in eq.\ (\ref{eq:rho}) assuming that 
each of the isolated leads is in the thermal equilibrium 
\begin{equation}
\widetilde{\rho}(-\infty)
\ =  \ 
  { e^{-\beta[H_1 - \mu_L N_L  -\mu_R N_R ]} 
    \over
       \mbox{Tr} \   
                 e^{-\beta[H_1 - \mu_L N_L  -\mu_R N_R ]}
   } 
    \;, 
\label{eq:rho_ini}
\end{equation}
where $N_{\lambda} = \sum_{k\sigma} 
c^{\dagger}_{k \lambda \sigma} c^{\phantom{\dagger}}_{k \lambda \sigma}$
for $\lambda=L,\, R$.
Then the expectation value of ${\cal O}_H(t)$ with 
respect to $\widehat{\rho}(0)$ can be written in the form of  
a time-ordered function along the Keldysh contour,   
\begin{align}
& \langle {\cal O}_H(t) \rangle \ \equiv \ 
            \mbox{Tr} \left[\,\widehat{\rho}(0) \,{\cal O}_H(t) \, \right]  
\nonumber
\\    
  &= \ 
  \text{Tr}
       \left[\, \widetilde{\rho}(-\infty) \,
      {\cal U}(-\infty,+\infty) \, 
      {\cal U}(+\infty,t) \, \widetilde{\cal O}(t) 
      \, {\cal U}(t,-\infty) \,\right]
.
\label{eq:average_H}
\end{align}
Note that ${\cal U}(-\infty,+\infty) = {\cal U}^{\dagger}(+\infty,-\infty)$ 
can be expressed in terms of  
the anti-time-ordering operator $\widetilde{\text{T}}$ as 
\begin{equation}
{\cal U}(-\infty,+\infty) =
 \widetilde{\text{T}} \exp 
\left[\, \text{i} \int_{-\infty}^{\infty} \text{d}t' 
\, \widetilde{H}_2 (t') \,\right] 
\label{eq:U_2_tild}\;.
\end{equation}

\section{Matrix Dyson Equation} 
\label{sec:Matrix_Dyson}

In the matrix notation used in eq.\ (\ref{eq:Keldysh_Matrix}),
the Dyson equation for the intra-site Green's function 
at the dot $\mbox{\boldmath $G$}(\omega)$ and that for
the inter-site ones between the dot and leads are given by
\begin{align}
& \mbox{\boldmath $G$}(\omega) \ = \,
    \mbox{\boldmath $g$}(\omega)  
 + \mbox{\boldmath $g$}(\omega)\,
\mbox{\boldmath $\Sigma$}(\omega)\,
\mbox{\boldmath $G$}(\omega)         
  \nonumber \\
 & \qquad \qquad  
 + v_L \, \mbox{\boldmath $g$}(\omega)\,
\mbox{\boldmath$\tau$}_3\,
\mbox{\boldmath $G$}_{Ld}(\omega) 
+ v_R \, \mbox{\boldmath $g$}(\omega)\,
\mbox{\boldmath$\tau$}_3\,
\mbox{\boldmath $G$}_{Rd}(\omega) 
\label{eq:Dyson_free_00}
\;,
\\
    &\mbox{\boldmath $G$}_{\lambda d}(\omega) \, = \,
v_{\lambda}^{\phantom{0}} \, 
\mbox{\boldmath $g$}_{\lambda}(\omega)\,
\mbox{\boldmath$\tau$}_3\,
\mbox{\boldmath $G$}(\omega) 
\label{eq:Dyson_free_-10}
         \;, \\
   & \mbox{\boldmath $G$}_{d \lambda}(\omega)  \, = \, 
v_{\lambda}^{\phantom{0}} \, 
\mbox{\boldmath $G$}(\omega)\, 
\mbox{\boldmath$\tau$}_3\,
\mbox{\boldmath $g$}_{\lambda}(\omega) 
\label{eq:Dyson_free_0-1} 
\;, \\
& \mbox{\boldmath$\tau$}_3 \, = \, 
\left[ 
          \begin{matrix}
                 1 & \phantom{-}0  \cr   
                    0 & -1 \cr 
          \end{matrix}
                                \right] 
         \;,
\qquad 
\mbox{\boldmath $P$} \, = \, {1 \over \sqrt{2} }
\left[ 
          \begin{matrix} 
                    \phantom{-}1 & 1 \cr   
                              -1 & 1 \cr  
          \end{matrix} 
\right] 
\;.
\end{align}
Here the elements of 
    $\mbox{\boldmath $G$}_{\lambda d}(t)$ and
    $\mbox{\boldmath $G$}_{\lambda d}(t)$ 
are defined by 
 $G^{--}_{\lambda d}(t) = - \text{i} \,
\langle \text{T}\,
c^{\phantom{\dagger}}_{\lambda\sigma}(t)\, d^{\dagger}_{\sigma}(0)
\rangle$, $\ldots$, 
and  $G^{--}_{d \lambda}(t) = - \text{i} \,
\langle \text{T}\,
d^{\phantom{\dagger}}_{\sigma}(t)\, c^{\dagger}_{\lambda\sigma}(0)
\rangle $, $\ldots$, respectively. 
The Green's functions for 
the isolated leads $\mbox{\boldmath $g$}_{\lambda}(\omega)$ 
include the distribution functions for the initial state,  
\begin{align}
\mbox{\boldmath $g$}_{\lambda}(\omega) \,&= 
\mbox{\boldmath $P$}
\left[ 
          \begin{matrix}
                    0 & g_{\lambda}^a(\omega)  \cr   
                    g_{\lambda}^r(\omega) & g^K_{\lambda}(\omega) \cr  
          \end{matrix}
                                \right] 
 \mbox{\boldmath $P$}^{-1}
 , 
\label{eq:gL_gR}
\\
g^r_{\lambda}(\omega) 
\,&= \,\sum_k 
{
\left|\phi_{k \lambda}^{\phantom{0}}(r_\lambda^{\phantom{0}})\right|^2
\over 
\omega - \epsilon_{k\lambda} + \text{i}\,0^+} 
\;,
\label{eq:g_lead}
\\
g^K_{\lambda}(\omega) \,&\equiv \,
[ 1 - 2\, f_{\lambda}(\omega) ]  
  [\, g_{\lambda}^{r}(\omega) -  g_{\lambda}^{a}(\omega) \,] 
\label{eq:F0_11}
   \;.
\end{align}
For the isolated dot,  
the Green's function can be written as  
 $\left\{ \mbox{\boldmath $g$}(\omega) \right\}^{-1} 
 =  \left(\omega -E_d\right)  \mbox{\boldmath$\tau$}_3$.
Note that 
the contribution of the element $g^K(\omega)$ which 
corresponds to eq.\ (\ref{eq:F0_11}) vanishes for the isolated dot 
because of a property 
$\{ g^{r}(\omega) \}^{-1}  
g^K(\omega)  \{ g^{a}(\omega) \}^{-1} =0$.\cite{HDW2} 
Using eqs.\ (\ref{eq:Dyson_free_00}) and (\ref{eq:Dyson_free_-10}), 
we have 
\begin{align}
   &\left\{ \mbox{\boldmath $G$}(\omega) \right\}^{-1} 
    \ = \ 
    \left\{ \mbox{\boldmath $G$}_0(\omega) \right\}^{-1} 
        \ - \ 
\mbox{\boldmath $\Sigma$}(\omega)\;, 
\label{eq:Dyson_onsite}
\\
&\left\{ \mbox{\boldmath $G$}_0(\omega) \right\}^{-1} 
    \, \equiv \ 
    \left\{ \mbox{\boldmath $g$}(\omega) \right\}^{-1} 
        \ - \ 
\mbox{\boldmath $\sigma$}(\omega)\;,
\label{eq:Dyson_onsite_free}
\\
& \; 
\mbox{\boldmath $\sigma$}(\omega) 
\  \equiv \ 
v_L^2\, 
\mbox{\boldmath$\tau$}_3\,
\mbox{\boldmath $g$}_{L}(\omega)\, 
\mbox{\boldmath$\tau$}_3
\, + \, 
v_R^2\, 
\mbox{\boldmath$\tau$}_3\,
\mbox{\boldmath $g$}_{R}(\omega)\, 
\mbox{\boldmath$\tau$}_3
\;.
\label{eq:self_U=0}
\end{align}
Then, 
using eqs.\ (\ref{eq:gL_gR})--(\ref{eq:F0_11}), 
$\mbox{\boldmath $\sigma$}(\omega)$  is written in the form
\begin{align}
 \mbox{\boldmath $\sigma$}(\omega)
 &\,= \,
 \mbox{\boldmath $P$} 
\left[ 
 \begin{matrix}
  \Omega_0(\omega) & \sigma^{r}(\omega)   \cr
           \sigma^{a}(\omega) & 0  \cr  
  \end{matrix}         
                             \right] 
 \mbox{\boldmath $P$}^{-1}
\;,
\\
\sigma^{r}(\omega) &\,=\, v_L^2\, g_{L}^r(\omega) 
+ v_R^2\, g_{R}^r(\omega) \;,
\label{eq:self_U=0_ra} 
\\
\Omega_0(\omega) &\,=\, 
    v_L^2\,g^K_{L}(\omega) + v_R^2\,g^K_{R}(\omega) 
    \;.
\label{eq:self_U=0_omega}
\end{align}
When the density of states $\text{Im}\, g_{\lambda}^r(\omega)$ 
 is a constant and its band width is infinity,  
$\sigma^{r}(\omega)$ becomes pure imaginary as 
 $v_{\lambda}^2 \,g^{r}_{\lambda}(\omega) =  - \text{i}\, \Gamma_{\lambda}$. 
We obtain the explicit form of $\mbox{\boldmath $G$}_0(\omega)$ 
given by eqs.\ (\ref{eq:G_0^--})--(\ref{eq:G_0^++}) 
via eq.\ (\ref{eq:Dyson_onsite_free}). 
Through eq.\ (\ref{eq:Dyson_onsite}),   
$\mbox{\boldmath $G$}(\omega)$ can be expressed in similar forms; 
\begin{align}
 &\!
 G^{--}(\omega) =  
 \bigl[ 1-\widetilde{f}_{\text{eff}}(\omega) \bigr]
     G^r(\omega)
      +   \widetilde{f}_{\text{eff}}(\omega)\,G^a(\omega) 
,
\label{eq:G_^--} 
\\
&\!
G^{-+}(\omega) =  
-\,\widetilde{f}_{\text{eff}}(\omega)
        \left[\,G^r(\omega)- G^a(\omega)\,\right] , 
\label{eq:G_^-+}
\\
&\!
G^{+-}(\omega) =  
\bigl[ 1-\widetilde{f}_{\text{eff}}(\omega) \bigr]
        \left[\,G^r(\omega)- G^a(\omega)\,\right] , 
\label{eq:G_^+-}
\\
 & \!
 G^{++}(\omega) =  
 -\bigl[ 1-\widetilde{f}_{\text{eff}}(\omega) \bigr]
     G^a(\omega)
      -   \widetilde{f}_{\text{eff}}(\omega) \,G^r(\omega),
\label{eq:G_^++}
\end{align}
where $G^r(\omega)$ and 
$\widetilde{f}_{\text{eff}}(\omega)$ are given by 
\begin{align}
 G^{r}(\omega) \,&= \,{1 \over \omega - E_d 
  -\sigma^r(\omega) -\Sigma^r(\omega)} \;,
 \label{eq:Gr_00}
\\
\widetilde{f}_{\text{eff}}(\omega) & \equiv
{ f_L(\omega) \,\Gamma_L 
  + f_R(\omega) \,\Gamma_R - 
  {\displaystyle 1\over \displaystyle 2\text{i}}\,\Sigma^{-+}(\omega)
  \over 
 \Gamma_L +\Gamma_R - \text{Im} \Sigma^r(\omega)} \;.
\label{eq:f_int}
\end{align}
Here $\widetilde{f}_{\text{eff}}(\omega)$ is real 
because $\Sigma^{-+}(\omega)$ is pure imaginary,
and at equilibrium it becomes the usual Fermi function 
since $\left.\Sigma^{-+}(\omega)\right|_{V=0} = 
 2\text{i}\,f(\omega) \,
 \text{Im}\left.\Sigma^{r}(\omega)\right|_{V=0}$.\cite{HDW2}
Note that $\Sigma^r = \Sigma^{--} + \Sigma^{-+}$.
There are some additional relations:
$\Sigma^{a}(\omega)= \left\{\Sigma^{r}(\omega)\right\}^*$, 
and $\Sigma^{--}(\omega)= - \left\{\Sigma^{++}(\omega)\right\}^*$.


\section{Current and Charge} 
\label{sec:current}

The operator for the current flowing 
from the left lead to the dot $I_L$ 
and that from the dot to the right lead $I_R$ are given by 
\begin{equation}
I_{\lambda} \, = \, \text{i}\,{\sl e}\,w_{\lambda}  \sum_{\sigma}\,  
v_{\lambda}^{\phantom{0}} \left(\,  
c^{\dagger}_{\lambda \sigma} d^{\phantom{\dagger}}_{\sigma} 
-d^{\dagger}_{\sigma} c^{\phantom{\dagger}}_{\lambda \sigma}\,\right) ,
\end{equation}
where $w_{R} = 1$ and $w_{L}=-1$. 
The equation of continuity 
is written as $\partial n_d /\partial t + I_R -I_L =0$ with  
 $n_d=\sum_{\sigma} n_{d\sigma}$. 
The averages of these operators with respect to $\widehat{\rho}(0)$ are
\begin{align}
\langle n_d  \rangle 
\, &= \, 
 2 \int_{-\infty}^{\infty} {\text{d}\omega \over 2\pi \text{i}}
 \,  G^{-+}(\omega) \;,  
\\
\langle I_{\lambda} \rangle 
\, &= \,  2 {\sl e}  \,w_{\lambda} 
\int_{-\infty}^{\infty} {\text{d}\omega \over 2\pi}
\, v_{\lambda}^{\phantom{0}} 
\left[\, G_{d \lambda}^{-+}(\omega) -  G_{\lambda d}^{-+}(\omega) \,\right] .
\label{eq:current_green}
\end{align}
Note that $G_{\lambda d}^{-+}$ and  $G_{d \lambda}^{-+}$ can be 
expressed in terms of $\mbox{\boldmath $G$}$ using 
eqs.\ (\ref{eq:Dyson_free_-10})--(\ref{eq:Dyson_free_0-1}).
Specifically, if the couplings with the leads satisfy a condition 
 $\Gamma_L(\omega) \propto \Gamma_R(\omega)$, 
the expectation value of the current can be rewritten as\cite{MW} 
\begin{align}
 &\!\!\!\!
 I\, = 
\frac{ 
\Gamma_L\langle\, I_R\, \rangle 
+ \Gamma_R\langle\, I_L\, \rangle}{\Gamma_R + \Gamma_L} 
\\
 & 
  =\,  {2 {\sl e} \over h}  \int_{-\infty}^{\infty} 
 \! \text{d}\omega 
   \left[\, f_L - f_R \, \right] 
\frac{4\, \Gamma_L \Gamma_R}{\Gamma_R + \Gamma_L} 
 \left[\, - {\rm Im}\, G^r(\omega) \,\right] .
\label{eq:caroli}
\end{align}
Note that $\langle I_L \rangle =\langle I_R \rangle$ in 
the stationary state.


\begin{thebibliography}{}



\bibitem{Goldharber}
D. Goldharber-Gordon, {\em et al.}:
Nature {\bf 391} (1998) 156.

\bibitem{Cronenwett}
S. M. Cronenwett {\em et al.}: 
Science {\bf 281} (1998) 540.

\bibitem{Simmel}
F. Simmel {\em et al.}:
Phys.\ Rev.\ Lett.\ {\bf 83} (1999) 804.

\bibitem{VanDerWiel}
W. G. van der Wiel, {\em et al.}:
Science {\bf 289} (2000) 2105.




\bibitem{NgLee}
T. K. Ng and P. A. Lee: Phys.\ Rev.\ Lett.\ {\bf 61} (1988) 1768.

\bibitem{GlatzmanRaikh}
L. I. Glazman and M. E. Raikh: 
JETP Lett.\ {\bf 47} (1988) 452.

\bibitem{Kawabata}
A. Kawabata: J.\ Phys.\ Soc.\ Jpn.\ {\bf 60} (1991) 3222.


\bibitem{Hewson}
A.\ C.\ Hewson: 
{ \em  The Kondo Problem to Heavy Fermions\/} 
(Cambridge University Press, Cambridge, 1993). 


\bibitem{Izumida}
W. Izumida, O. Sakai, and S. Suzuki: 
J.\ Phys.\ Soc.\ Jpn.\ {\bf 70} (2001) 1045.



\bibitem{WM} 
N.\ S.\ Wingreen and Y.\ Meir:
Phys. Rev. {\bf B 49} (1994)  11040.


\bibitem{HDW2}
S. Hersfield, J. H. Davies, and J. W. Wilkins: 
Phys.\ Rev.\ B {\bf 46} (1992) 7046.


\bibitem{TheoriesII}
A. Yeyati  {\em et al.}: 
Phys.\ Rev.\ Lett.\ {\bf 71} (1993) 2991;
M. H. Hettler {\em et al.}:  
Phys.\ Rev.\ Lett.\ {\bf 73} (1994) 1967;
A. Schiller and S. Hershfield: 
Phys.\ Rev.\ B {\bf 58} (1998) 14978;
H. Schoeller and J. K\"onig: 
Phys.\ Rev.\ Lett. {\bf 84} (2000) 3686;
P. Coleman {\em et al.}: 
cond-mat/0108001; Phys.\ Rev.\ Lett.\ {\bf 86} (2001) 4088;
A. Rosch, {\em et al.}: Phys.\ Rev.\ Lett.\ {\bf 87} (2001) 156802.


\bibitem{KNG}
A. Kaminski, Yu. V. Nazarov, and L. I. Glazman:
  Phys.\ Rev.\ B {\bf 62} (2000) 8154.


\bibitem{ao10}
A.\ Oguri: Phys.\ Rev.\ B {\bf 64} (2001) 153305.


\bibitem{Nozieres}
P.\ Nozi\`{e}res: J.\ Low Temp.\ Phys.\ {\bf 17} (1974) 31.

\bibitem{Yamada}
K.\ Yamada: Prog.\ Theor.\ Phys.\  {\bf 53} (1975) 970;
Prog.\ Theor.\ Phys.\  {\bf 54} (1975) 316. 



\bibitem{LuttingerWard} 
J. M. Luttinger and J. C. Ward: Phys.\ Rev.\  {\bf 118} (1960) 1417.





\bibitem{KeldyshCaroliLandau} 
L. V. Keldysh: 
Sov. Phys. JETP {\bf 20} (1965) 1017;
C.\ Caroli, {\em el al.}: J.  Phys. C {\bf 4} (1971) 916.
We have used the notation given by   
E. M. Lifshitz and L. P. Pitaevskii,
{\em  Physical Kinetics\/} 
(Pergamon Press, Oxford, 1981).




\bibitem{PathIntegral}
See, for instance,
K. C. Chou {\em et al.}: Phys. Rep. {\bf 118} (1985) 1;
J. W. Negele and H. Orland: 
{\em Quantum Many-Particle Systems \/} (Addison-Wesley, Redwood City, 1987).



\bibitem{ZlaticHorvatic}
V. Zlati\'c and B. Horvati\'c: Phys.\ Rev.\ B {\bf 28} (1983) 6904.



\bibitem{HorvaticSokcevicZlatic} 
B. Horvati\'c, D. \v{S}ok\v{c}evi\'c, and V. Zlati\'c: 
  Phys.\ Rev.\ B {\bf 36} (1987) 365.



\bibitem{Logan} 
D. E. Logan and N. L. Dickens:
J. Phys.: Condes.\ Matter {\bf 14} (2002) 3605.


\bibitem{HewsonADV} 
 A.\ C.\ Hewson: Adv.\ Phys.\ {\bf 43} (1994) 543.


\bibitem{FermiLiquidRelations}
$\widetilde{\gamma} \equiv 
1  -  {\partial \Sigma^r(\omega) / \partial \omega }
|_{\omega=0}$ and 
$\widetilde{\chi}_{s}  = 
\widetilde{\chi}_{\uparrow\uparrow} 
- \widetilde{\chi}_{\uparrow\downarrow} 
$ with 
$\widetilde{\chi}_{\sigma\sigma'} 
\equiv \delta_{\sigma\sigma'}  -  
{\partial \Sigma_{\sigma}^r(0) / \partial h_{\sigma'} }
|_{h_{\sigma'}=0}
$,
where $h_{\sigma'}$ is an external field described by 
$H_{\text{ex}} 
= -  \Sigma_{\sigma} h_{\sigma}n_{d\sigma}
$. 
Note 
that $\widetilde{\gamma} = \widetilde{\chi}_{\uparrow\uparrow}$.
\cite{Yamada}





\bibitem{KawakamiOkiji} 
N. Kawakami and A. Okiji:
Solid State Commun. {\bf 43} (1982) 467. 

\bibitem{WiegmanTsvelick} 
B. Wiegman and A. M. Tsvelick:
J. Phys. C {\bf 16} (1983) 2281.



 
\bibitem{MW} 
Y.\ Meir and N.\ S.\ Wingreen: 
Phys. Rev. Lett. {\bf 68} (1992)  2512.





\end{thebibliography}
\end{document}